\documentclass[twocolumn,aps,graphicx,floatfix]{revtex4}
\usepackage[dvips]{graphicx}

\newcommand{\mypicwidth}{0.78\columnwidth}
\newcommand{\myext}{ps}
\newcommand{\n}[2]{n_{#1}^{#2}}
\newcommand{\nv}[2]{{\bf n}_{{\rm v}#1}^{#2}}

\newcommand{\w}[2]{w_{#1}^{#2}}
\newcommand{\wv}[2]{{\bf w}_{{{\rm v}#1}}^{#2}}

\newcommand{\wm}[2]{{\bf \hat{w}}_{{{\rm m}#1}}^{#2}}
\newcommand{\mymat}[1]{{\hat{\mathbf{#1}}}}

\newcommand{\nmtl}[2]{{\mymat{n}_{{\rm m}#1}}^{#2}}

\newcommand{\sphere}{S}
\newcommand{\colloid}{C}
\newcommand{\needle}{N}
\newcommand{\polymer}{P}

\newcommand{\sphereneedle}{{\rm\it S\hspace{-1pt}N}}

\newcommand{\rn}{\rho_{N}^{\ast}}

\newcommand{\epr}{\eta_{\polymer}^r}
\newcommand{\ep}{\eta_{\polymer}}
\newcommand{\ec}{\eta_{\colloid}}

\newcommand{\rhon}{\rho_{\needle}}

\newcommand{\Fexc}{F_{\rm exc}}

\newcommand{\kb}{k_{\rm B}}
\newcommand{\kt}{k_{\rm B} T}

\newcommand{\venergy}[1]{V_{#1}}
\newcommand{\phifmt}[1]{\Phi_{#1}}

\newcommand{\rv}{{\bf r}}

\newcommand{\xv}{{\bf x}}

\newcommand{\Ov}{{\bf \Omega}}

\newcommand{\fumindex}{\nu}

\begin{document}
\author{Matthias Schmidt\footnote{
permanent address: Institut f{\"u}r Theoretische Physik II,
  Heinrich-Heine-Universit{\"a}t D{\"u}sseldorf,
  Universit{\"a}tsstra{\ss}e 1, D-40225 D{\"u}sseldorf, Germany.
}, Alan R. Denton}
\title{
Colloids, polymers, and needles:
Demixing phase behavior
}
\date{4 October 2001}
\address{
Department of Physics,
North Dakota State University,
Fargo, ND 58105-5566, USA.
}

\begin{abstract}
We consider a ternary mixture of hard colloidal spheres, ideal polymer
spheres, and rigid vanishingly thin needles, which model stretched
polymers or colloidal rods.  For this model we develop a
geometry-based density functional theory, apply it to bulk fluid
phases, and predict demixing phase behavior. In the case of no
polymer-needle interactions, two-phase coexistence between
colloid-rich and -poor phases is found.  For hard needle-polymer
interactions we predict rich phase diagrams, exhibiting three-phase
coexistence, and reentrant demixing behavior.
\end{abstract}

\maketitle

\newpage
\section{Introduction}
The richness of phase behavior of systems with purely repulsive
interactions depends crucially on the number of components. For a
one-component system like colloidal hard spheres, there occurs a
freezing transition from a single fluid phase to a dense
crystal. Adding a second component, such as non-adsorbing globular
polymer coils\cite{asakura54}, or rod-like
particles\cite{vliegenthart99,kluijtmans00} generates an effective
depletion-induced attraction between colloidal spheres, leading to the
possibility of demixing. This transition is an analog of the
vapor-liquid transition in simple fluids: The phase that is
concentrated in one of the components corresponds to a liquid, while
the dilute phase corresponds to a vapor, and one frequently refers to
such phases as colloidal liquid and colloidal vapor, although the
``vapor'' is concentrated in the added component.

Generic theoretical models for such systems are those introduced by
Asakura and Oosawa (AO) and independently by
Vrij\cite{asakura54,vrij76}, Bolhuis and Frenkel (BF)\cite{bolhuis94},
and Widom and Rowlinson (WR)\cite{widom70}. The AO model comprises
hard colloidal spheres mixed with polymer spheres that are ideal
amongst themselves but cannot penetrate the colloids. The BF model
adds stiff vanishingly thin needles to a hard sphere system. Because
of their vanishing thickness, the needles do not interact with one
another. Clearly, both models are similar in spirit, as a
non-interacting component is added to hard spheres.  In the WR model
this is different; two species of spheres interact symmetrically, such
that hard core repulsion occurs only between particles of unlike
species. Hence a pure system of either component is an ideal gas.  All
of these model binary mixtures exhibit liquid-vapor phase separation,
well-established by computer simulations and theories
\cite{gast83,lekkerkerker92,dijkstra99,louis99,dijkstra00I,brader01phd}.
The WR model\cite{widom70,guerrero76,rowlinson80,rowlinson82} has been
studied with a range of approaches, including mean-field theory
(MFT)\cite{rowlinson82}, Percus-Yevick (PY) integral equation
theory\cite{rowlinson80,shew96,yethiraj00}, scaled-particle theory
(SPT) \cite{bergmann76}, as well as computer
simulations\cite{shew96,johnson97,borgelt90}.  The precise location of
the liquid-vapor critical point was located by simulations about 50
percent higher than previously thought\cite{shew96,johnson97}, still a
challenge for theories (for a recent integral-equation closure, see
Ref.\ \cite{yethiraj00}).

In the AO and BF cases a reservoir description has proven to be
useful.  The reservoir density of either polymers or needles rules the
strength of effective attraction and hence plays a role similar to
(inverse) temperature in simple substances.  Although the WR model
features an intrinsic symmetry which seems to preclude such a
description, an effective model can also be
formulated\cite{rowlinson82}.  In the present work we consider the
phase behavior of a mixture of spheres, polymers and needles, a
natural combination of the above binary cases.  We note that our
ternary model may provide insight into certain real systems, such as
paints, which contain colloidal latex and pigment particles, polymer
thickeners and dispersants, as well as many other
components\cite{satas91}.

Density functional theory (DFT)\cite{evans92} is a powerful approach
to equilibrium statistical systems, possibly under influence of an
external potential. Building on Rosenfeld's work\cite{Rosenfeld89}, a
geometry-based approach was recently proposed that also predicts bulk
properties, without the need of any input, allowing the
AO\cite{schmidt00cip}, BF\cite{schmidt01rsf}, and WR\cite{schmidt01wr}
models to be treated.  Here we combine these tools to derive a DFT for
ternary systems.

In Sec.\ \ref{SECmodels} we define the model ternary mixtures of
spheres, polymers, and needles. In Sec.\ \ref{SECtheory} the DFT is
developed. Application to bulk phases in Sec.\ \ref{SECresults} yields
the phase behavior. We finish with concluding remarks in Sec.\
\ref{SECdiscussion}.

\section{The Model}
\label{SECmodels}
\begin{figure}
  \begin{center}
    \includegraphics[width=\columnwidth,angle=0]{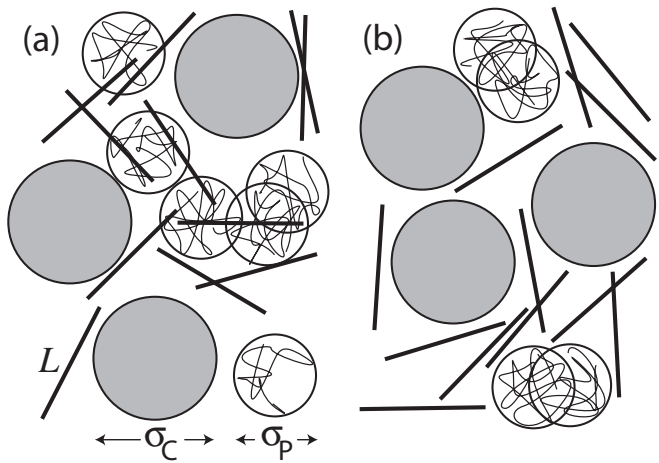}
    \caption{ Sketch of the ternary mixture of colloidal hard spheres
    with diameter $\sigma_C$, ideal polymer spheres of diameter
    $\sigma_P$ and vanishingly thin needles of length $L$.  Different
    cases for interactions between polymers and needles are depicted:
    a) no interactions; b) excluded volume interactions.}
    \label{FIGmodel} \end{center}
\end{figure}

We consider a mixture of colloidal hard spheres (species $C$) of
radius $R_C$, globular polymers (species $P$) of radius $R_P$, and
vanishingly thin needles (species $N$) of length $L$, with respective
number densities $\rho_C(\rv)$, $\rho_P(\rv)$ and $\rho_N(\rv,\Ov)$,
where $\rv$ is the spatial coordinate and $\Ov$ is a unit vector
pointing along the needle axis (see Fig.\ \ref{FIGmodel}).  The pair
interaction between colloids is $V_{CC}=\infty$ if the separation $r$
between sphere centers is less than $2R_C$, and zero otherwise. The
pair interactions between like particles of both other components
vanish for all distances: $V_{PP}=V_{NN}=0$.  For polymers this is an
assumption strictly valid only at the theta point; for needles it
becomes exact in the present limit of large aspect ratio, where
overlapping needles contribute a negligible fraction of
configurations.  The colloidal spheres interact with both other
components via excluded volume: The pair interaction between colloids
and polymers is $V_{CP}=\infty$ if $r<R_C+R_P$, and zero otherwise;
the interaction between colloids and needles is $V_{CN}=\infty$, if
both overlap, and zero otherwise.  What remains to be prescribed is
the interaction between needles and polymers. We consider two cases:
i) ideal interactions such that $V_{PN}=0$ for all
 distances, and ii)
excluded volume interactions such that $V_{PN}=\infty$ if needle and
polymer overlap, and zero otherwise.  We denote the sphere diameters
by $\sigma_C=2R_C$, $\sigma_P=2R_P$, the sphere packing fractions by
$\eta_C=4\pi R_C^3 \rho_C/3$, $\eta_P=4\pi R_P^3 \rho_P/3$, and use a
dimensionless needle density $\rn=\rho_N L^3.$

\section{Density functional theory}
\label{SECtheory}
\subsection{Weight functions}
We start with a geometrical representation of the particles in terms
of weight functions $\w{\mu}{i}$, where $\mu=3,2,1,0$ corresponds to
the particles' volume, surface, integral mean curvature and Euler
characteristic, respectively\cite{rosenfeld94convex}, and $i=C,P,N$
labels the species. We will use $S$ as a unifying symbol for the
spherical species $C$ and $P$, and denote the radius as $R$, where
$R=R_C,R_P$ for $S=C,P$, respectively. The weight functions are
determined to give the hard core Mayer bonds
$f_{ij}=\exp(\venergy{ij})-1$ by a linear combination of terms
$\w{\gamma}{i}(\rv) * \w{3-\gamma}{j}(\rv)$, where the star denotes
the convolution, $g(\rv)*h(\rv)=\int {\rm d}^3 x \, g(\xv)
h(\rv-\xv)$.

For spheres, the usual weight functions\cite{Rosenfeld89,tarazona00}
are
\begin{eqnarray}
\w{3}{\sphere}( \rv ) =& \theta( R - r),  \hspace{3mm}
\w{2}{\sphere}( \rv ) &= \delta( R - r), \label{EQwsfirst}\\
\wv{2}{\sphere}( \rv ) =& \w{2}{\sphere}(\rv) \, \rv/r,  \hspace{3mm}
\wm{2}{\sphere}( \rv ) &=  \w{2}{\sphere}(\rv) [ \rv\rv/r^2 - \mymat{1}/3],
\label{EQwssecond}
\end{eqnarray}
where $r=|\rv|$, $\delta(r)$ is the Dirac distribution, $\theta(r)$ is
the step function, and $\mymat{1}$ is the identity matrix.  Further
linearly dependent weights are $\w{1}{\sphere}(\rv) =
\w{2}{\sphere}(\rv)/(4 \pi R), \wv{1}{\sphere}(\rv) =
\wv{2}{\sphere}(\rv)/(4 \pi R), \w{0}{\sphere}(\rv) =
\w{1}{\sphere}(\rv)/R$. Note that these weights have different
tensorial rank: $\w{0}{\sphere}$, $\w{1}{\sphere}$, $\w{2}{\sphere}$,
$\w{3}{\sphere}$ are scalars; $\wv{1}{\sphere}$, $\wv{2}{\sphere}$ are
vectors; $\wm{2}{\sphere}$ is a (traceless) matrix.  These functions
give the Mayer bond between pairs of spheres\cite{Rosenfeld89} through
$-f_{\sphere\sphere}/2= \w{3}{\sphere}*\w{0}{\sphere} +
\w{2}{\sphere}*\w{1}{\sphere} - \wv{2}{\sphere}*\wv{1}{\sphere}$.
However, they are not sufficient to recover the sphere-needle Mayer
bond\cite{rosenfeld94convex}.  This is achieved through
\begin{equation}
 \w{2}{\sphereneedle}(\rv,\Ov)= 2 | \wv{2}{\sphere}(\rv) \cdot \Ov |,
\label{EQwsn}
\end{equation}
which contains information about {\em both} species: it is
non-vanishing on the surface of a sphere with radius $R$, but this
surface is ``decorated'' with an $\Ov$-dependence.  Furthermore, for
needles, we follow\cite{rosenfeld94convex} to obtain
\begin{eqnarray}
\w{1}{\needle}(\rv,\Ov) &=& \frac{1}{4}\int_{-L/2}^{L/2} {\rm d} l \,
  \delta(\rv + \Ov \, l)\label{EQwneedleone},\\
\w{0}{\needle}(\rv,\Ov) &=& \frac{1}{2}\left[\delta(\rv + \Ov L/2) +  
  \delta(\rv-\Ov L/2)\right],\label{EQwneedlezero}
\end{eqnarray}
and $\rv$ is the needle center of mass.  The function $\w{1}{\needle}$
describes the linear extent of a needle, whereas $\w{0}{\needle}$ is
characteristic of its endpoints. For vanishingly thin needles, both
surface and volume vanish, and so do the corresponding weights,
$\w{2}{\needle}=\w{3}{\needle}=0$.  Technically, the Mayer bond is
generated through $ -f_{\sphereneedle}(\rv,\Ov) = \w{3}{\sphere}(\rv)
* \w{0}{\needle}(\rv,\Ov) + \w{2}{\sphereneedle}(\rv,\Ov) *
\w{1}{\needle}(\rv,\Ov)$, where $\rv$ is the difference vector
between sphere and needle position.

\subsection{Weighted densities}
The weight functions are used to smooth the possibly highly
inhomogeneous density profiles by convolutions,
\begin{eqnarray}
  \n{\fumindex}{C}(\rv) &=& \rho_C(\rv) * \w{\fumindex}{C}(\rv),
    \label{EQnfirst}\\
  \n{\fumindex}{P}(\rv) &=& \rho_P(\rv) * \w{\fumindex}{P}(\rv),\\
  \n{2}{CN}(\rv,\Ov) &=& 
       \rho_C(\rv) * \w{2}{CN}(\rv,\Ov),\\
  \n{2}{PN}(\rv,\Ov) &=& 
       \rho_P(\rv) * \w{2}{PN}(\rv,\Ov),\\
  \n{\tau}{\needle}(\rv,\Ov) &=& 
       \rhon(\rv,\Ov) * \w{\tau}{\needle}(\rv,\Ov),\label{EQnlast}
\end{eqnarray}
where $\nu=0,1,2,3,v1,v2,m2$, and $\tau=0,1$; $\rho_C(\rv)$,
$\rho_P(\rv)$ and $\rhon(\rv,\Ov)$ are the one-body density
distributions of spheres, polymers and needles, respectively.  Note
that $\n{\fumindex}{C}, \n{\fumindex}{P}, \n{\fumindex}{N}$ are
``pure'' weighted densities, involving only variables of either
species\cite{Rosenfeld89,rosenfeld94convex}.  In contrast, $\n{2}{CN}$
and $\n{2}{PN}$ are a convolution of the sphere densities with
orientation-dependent weight function, combining characteristics of
both species\cite{schmidt01rsf}.

\subsection{Free energy density}
The Helmholtz excess free energy is obtained by integrating over a
free energy density,
\begin{equation}
\Fexc [ \rho_C, \rho_P, \rhon ] = 
\kt \int {\rm d}^3 x \int \frac{{\rm d}^2 \Omega}{4\pi} \,
\phifmt{} \left( \{\n{\gamma}{i}\} \right),
\label{EQfexc}
\end{equation}
where $\kb$ is Boltzmann's constant, $T$ is temperature, and the
(local) reduced excess free energy density $\phifmt{}$ is a simple
function (not a functional) of the weighted densities $\n{\gamma}{i}$.
This leads to a dependence of $\phifmt{}$ on orientation and position.
The variable $\xv$ runs over
space\cite{Rosenfeld89,rosenfeld94convex}, and $\Ov$ over the unit
sphere\cite{schmidt01rsf}.

The functional form of $\phifmt{}$ is obtained by consideration of the
exact zero-dimensional excess free energy.  We obtain
\begin{equation}
 \Phi = \Phi_{CC} + \Phi_{CP} + \Phi_{CN} + \lambda \Phi_{PN},
\end{equation}
where in the case of ideal polymer-needle interaction $\lambda=0$, and
for hard polymer-needle interaction $\lambda=1$.  In the following,
the arguments of the weighted densities are suppressed in the
notation; see Eqs.\ (\ref{EQnfirst})-(\ref{EQnlast}) for the explicit
dependence on $\rv$ and $\Ov$.  The hard sphere contribution, being
equal to the pure HS case \cite{Rosenfeld89,tarazona00}, is
\begin{eqnarray}
\phifmt{CC} &=&
  -\n{0}{C} \ln (1 - \n{3}{C})+
  \left(  \n{1}{C}\,\n{2}{C} - 
  \nv{1}{C} \cdot \nv{2}{C} \right)/({1 -
  \n{3}{C}})\nonumber\\ 
& &+
\left[{{\left(\n{2}{C}\right)^3}/3} - 
  \n{2}{C}\,{\left(\nv{2}{C}\right)^2} +
  3\left(\nv{2}{C} \cdot \nmtl{2}{C} \cdot \nv{2}{C}
\right.\right.\nonumber\\ 
&&  \left.\left.\phantom{{\left(\n{2}{C}\right)^3}/3}\left.\hspace{-8mm}
          -\,3\det \nmtl{2}{C}\right)/2
      \right]\right/[8\pi(1-\n{3}{C})^2].\label{EQphic}
\end{eqnarray}
The contribution due to interactions between colloids and polymers is
the same as in the pure AO case\cite{schmidt00cip} and is given by
\begin{equation}
\phifmt{CP} = 
 \sum_\nu \frac{\partial \phifmt{CC}}{\partial \n{\nu}{C}} \n{\nu}{P}.
\end{equation}
The contribution due to interactions between colloids and
needles\cite{schmidt01rsf} is
\begin{equation}
\phifmt{CN} =
  -\n{0}{\needle} \ln (1 - \n{3}{C}) +
  \frac{\n{1}{\needle} \n{2}{CN}}{1 -
    \n{3}{C}}\label{EQphicn}.
\end{equation}
Note that the simultaneous presence of $\Phi_{CP}$ and $\Phi_{CN}$ in
$\Phi$ does not generate artificial interactions between $P$ and $N$.
For vanishing $PN$ pair potential one can derive these terms from
consideration of multi-cavity distributions like in the binary
$CP$\cite{schmidt00cip,brader01phd} and $CN$ cases\cite{schmidt01rsf}.
In order to model the WR type interaction between polymers and needles
in the presence of the colloidal spheres we use
\begin{equation}
\phifmt{PN} =
  \frac{ \n{0}{N} \n{3}{P} + \n{1}{N} \n{2}{PN}
  }{1-\n{3}{C}}\label{EQphipn}.
\end{equation}
This can be derived as follows.  The starting point is a functional
for binary hard spheres with added needles. Linearization in one of
the sphere densities (which becomes the polymer species) is performed
in the same way as linearization of binary hard spheres leads to the
$CP$ functional\cite{brader01phd}.  In the absence of colloids, we
obtain $\Phi=\Phi_{CN}= \n{0}{N} \n{3}{P} + \n{1}{N} \n{2}{PN}$. Then
the density functional then can be rewritten as $\Fexc=-\int d^3r \int
d^3r' \int d^2 \Omega \rho_P(\rv) f_{PN}(\rv;\rv',\Ov)
\rho_N(\rv,\Ov)/(4\pi)$.  This is precisely (a generalization to
needles of) the mean-field DFT for the WR model\cite{rowlinson82}.
Although this does not feature the exact 0d limit, as the
geometry-based DFT\cite{schmidt01wr} for WR {\em spheres} does, we
expect differences to be small.

\section{Results}
\label{SECresults}
\subsection{Bulk fluid phases}
For homogeneous density profiles, $\rho_i=\rm const$, the integrations
in Eqs.\ (\ref{EQnfirst})-(\ref{EQnlast}) can be carried out explicitly.
The hard sphere contribution is equal to the Percus-Yevick
compressibility (and scaled-particle) result, which is
\begin{equation}
  \Phi_{CC} = \frac{3 \eta_C
    [3 \eta_C (2-\eta_C) - 2(1-\eta_C)^2\ln(1-\eta_C)]}
  {8 \pi R_C^3 (1-\eta_C)^2}.
\end{equation}
The colloid-polymer contribution is equal to that predicted by free
volume theory\cite{lekkerkerker92}, and rederived by
DFT\cite{schmidt00cip} as
\begin{eqnarray}
  \Phi_{CP} &=& 
    \frac{\eta_P/(8\pi R_P^3)}{(1-\eta_C)^3}
    \left\{
    3 q \eta_C \left[ 6(1-\eta_C)^2 + 
     3q(2-\eta_C -\eta_C^2)
\right.\right.
\nonumber\\       
& &\left.\left. \hspace{-5mm}
     +2q^2(1+\eta_C+\eta_C^2)  \right] - 
     6(1-\eta_C)^3 \ln (1 - \eta_C)
    \right\},
\end{eqnarray}
where $q=\sigma_P/\sigma_C$.  The colloid-needle contribution equals
the perturbative (around a pure hard sphere fluid) treatment of
Ref.\cite{bolhuis94}, which can be shown to equal the result from
application of scaled-particle theory\cite{barker76}, and
DFT\cite{schmidt01rsf}, and is given by
\begin{equation}
  \Phi_{CN} = \rhon\left[
  -\ln(1-\eta_C)+\
   \frac{3 L}{4R_C} \, \frac{\ec}{1-\eta_C}
  \right].
\end{equation}
The WR type polymer-needle contribution is
\begin{equation}
  \Phi_{PN} = \left(1+\frac{3L}{4R_P}\right)\frac{\rho_N \eta_P}{1-\eta_C}.
\end{equation}
For completeness, the ideal free energy contribution is
\begin{equation}
  \Phi_{\rm id} = \sum_{i=C,P,N} \rho_i 
  [\ln(\rho_i \Lambda_i^3) -1],
\end{equation}
where the $\Lambda_i$ are (irrelevant) thermal wavelengths of species
$i$.  This puts us into a position to obtain the reduced total free
energy per volume $\Phi_{\rm tot}=\Phi_{\rm id}+\Phi$ of any given
fluid state characterized by the bulk densities and relative sizes of
the three components.

\subsection{Phase diagram}
The general conditions for phase coexistence are equality of the total
pressures $p_{\rm tot}$, and of the chemical potentials $\mu_i$ in the
coexisting phases.  Equality of temperature is trivial in hard-body
systems.  For phase equilibrium between phases I and II,
\begin{eqnarray}
p_{\rm tot}^{\rm I} &=& p_{\rm tot}^{\rm II}\\
\mu_i^{\rm I} &=& \mu_i^{\rm II}, \quad i=C,P,N.
\end{eqnarray}
These are four equations for six unknowns (two statepoints each
characterized by three densities). Hence two-phase coexistence regions
depend parametrically on two free parameters. For three-phase
equilibrium between phases I, II, and III
\begin{eqnarray}
p_{\rm tot}^{\rm I} &=& p_{\rm tot}^{\rm II} = p_{\rm tot}^{\rm III}\\
\mu_i^{\rm I} &=& \mu_i^{\rm II} = \mu_i^{\rm III}, \quad i=C,P,N.
\end{eqnarray}
Eight equations for nine variables leave one free parameter. 

In our case $p_{\rm tot}/k_BT=-\Phi_{\rm tot}+\sum_{i=C,P,N} \rho_i \partial
\Phi_{\rm tot}/\partial \rho_i$, and $\mu_i=k_B T \partial \Phi_{\rm
tot}/\partial \rho_i$ yield analytical expressions. We solve the
resulting sets of equations numerically, which is straightforward.

\subsubsection{Ideal polymer-needle interaction}
Let us first explain our representation of the ternary phase
diagrams. We take the system densities $\ec,\ep,\rn$ as basic
variables. For given particle sizes, these span a three-dimensional
(3d) phase space. Each point in this space corresponds to a possible
bulk state, at some pressure $p_{\rm tot}$. Two-phase coexistence is
indicated by a pair of points that are joined by a straight tie
line. Accordingly, three phase coexistence is a triplet of points,
defining a triangle.  In order to graphically represent the phase
diagram, we show surfaces defined by one thermodynamic parameter being
constant. Such surfaces are conveniently taken such that coexistence
lines (and triangles) lie completely within the surface. Clearly, this
can be accomodated by imposing a constant value of $p_{\rm tot}$ or
any of $\mu_C,\mu_P$, and $\mu_N$. Here we choose $\mu_P=\rm const$,
and hence imagine controlling the system directly with $\ec$ and
$\rn$, but indirectly via coupling to a polymer reservoir of packing
fraction $\epr=(4\pi/3) (R_P/\Lambda_P)^3\exp(\mu_P/k_BT)$.  A
constant-$\epr$-surface is non-trivially embedded in the 3d phase
diagram.  To depict it graphically, we show projections onto the three
sides of the coordinate system, namely the $\ec-\rn$, $\ec-\ep$, and
$\ep-\rn$ planes, as well as a perspective 3d view.  Furthermore, we
indicate the accessible regions that are compatible with the
constraint of fixed $\epr$.  Their boundaries are implicitly defined
through $\epr(\ec=0,\ep,\rn)=\rm const$ and $\epr(\ec,\ep,\rn=0)=\rm
const$. Note that tielines are allowed to cross inaccessible regions.

\begin{figure}
  \begin{center}

    \includegraphics[width=\mypicwidth,angle=-90]{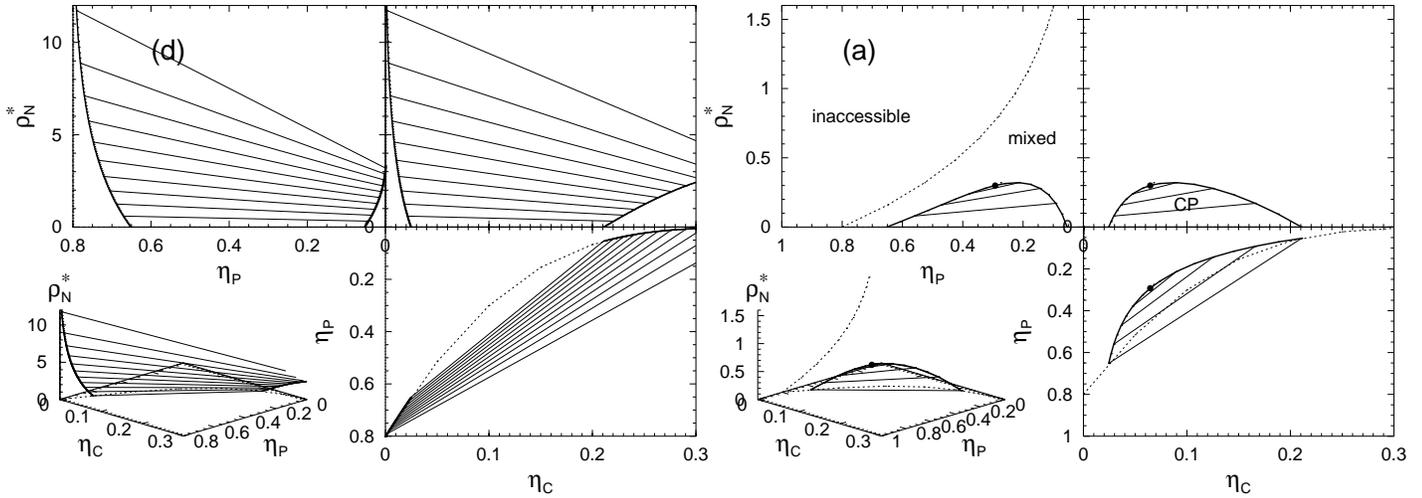}

    \caption{ Demixing phase diagram of a ternary
    colloid-polymer-needle mixture with ideal polymer-needle
    interactions for $\sigma_C=\sigma_P=L$, and $\eta_P^r=$0 (a), 0.5
    (b), 0.63831 (c), 0.8 (d).  Shown are binodals (lines), tielines
    between coexisting phases (thin lines), and critical points
    (dots).  }  \label{FIGid1} \end{center}
\end{figure}

For simplicity, and to establish a reference case, we initially ignore
polymer-needle interactions and consider equal particle sizes,
$\sigma_C=\sigma_P=L$. In the absence of polymer ($\epr=0$), colloids
and needles demix, as shown in Fig.\ \ref{FIGid1}a. Increasing the
packing fraction of polymers in the reservoir causes the demixed
region to grow and to shift to smaller $\ec$ and $\rn$ (see Fig.\
\ref{FIGid1}b for $\ep=0.5$). This behavior can be understood if
addition of a second depleting species simply enhances the
depletion-induced attraction between colloids.  Increasing $\epr$
further causes the critical point to hit the $\rn=0$ axis. This is
precisely the demixing critical point of the binary $CP$ (AO) model,
which is located at $\epr=0.63831$ (see Fig.\ \ref{FIGid1}c). Computer
simulations are currently being carried out to test the accuracy of
this value\cite{dijkstra01private}. For still larger $\epr$, the mixed
states become disconnected, hence there is no path between
colloid-rich and colloid-poor phases that does not pass through a
first-order phase transition (see Fig.\ \ref{FIGid1}d for $\epr=0.8$).

\subsubsection{Hard polymer-needle interaction}
Turning on the excluded volume interaction between polymers and
needles allows the possibility of demixing between these
components. In the absence of colloids, the $PN$ mixture is of WR
type: Interactions between particles of like species vanish, while
unlike particles interact with a hard core repulsion. Our case is a
generalization to non-spherical particle shapes. In the mean-field
treatment this does not affect the phase diagram, as only the net
excluded volume enters into the theory. This robustness is also
present in our approach.

\begin{figure}
  \begin{center}

    \includegraphics[width=\mypicwidth,angle=-90]{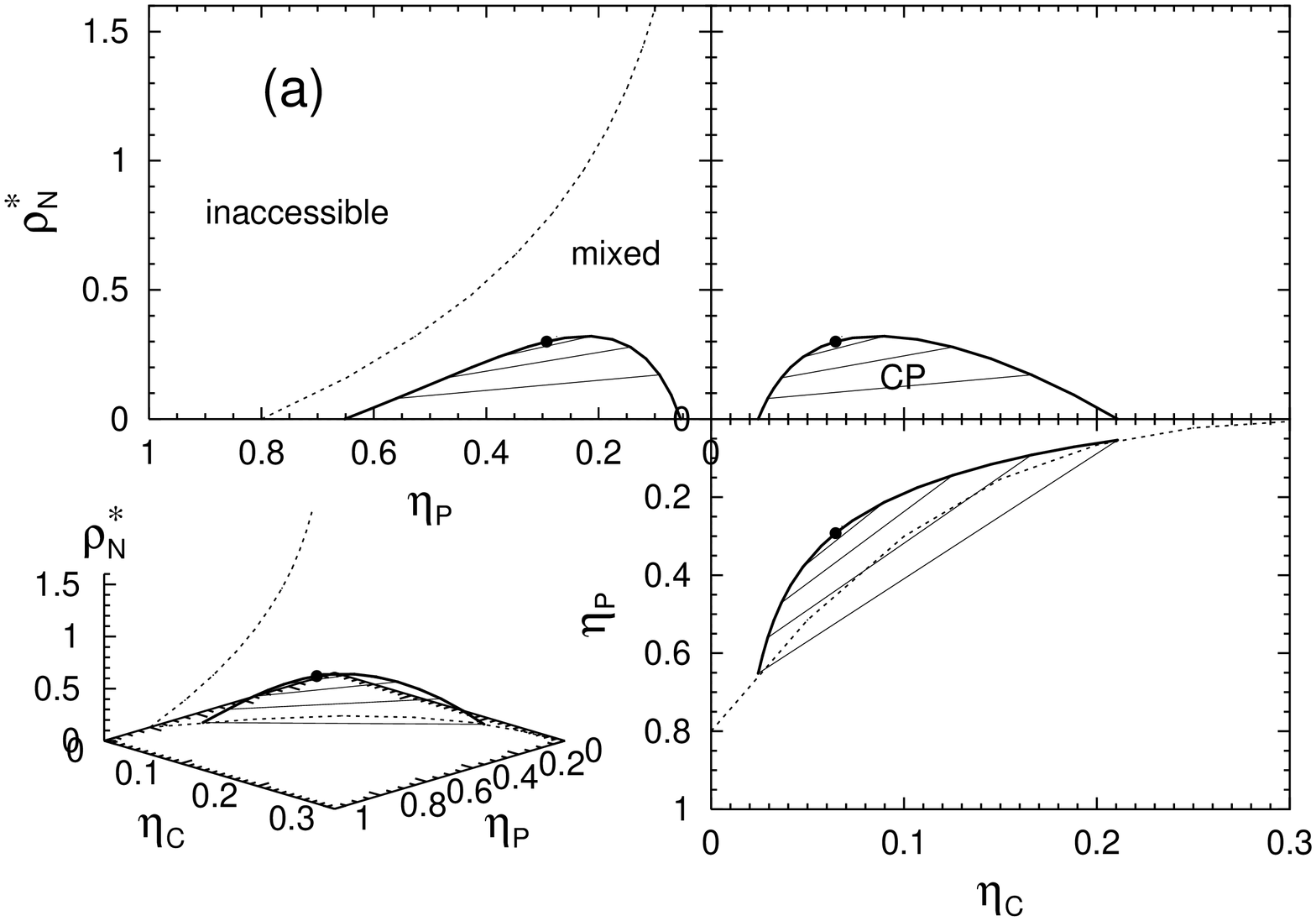}

    \includegraphics[width=\mypicwidth,angle=-90]{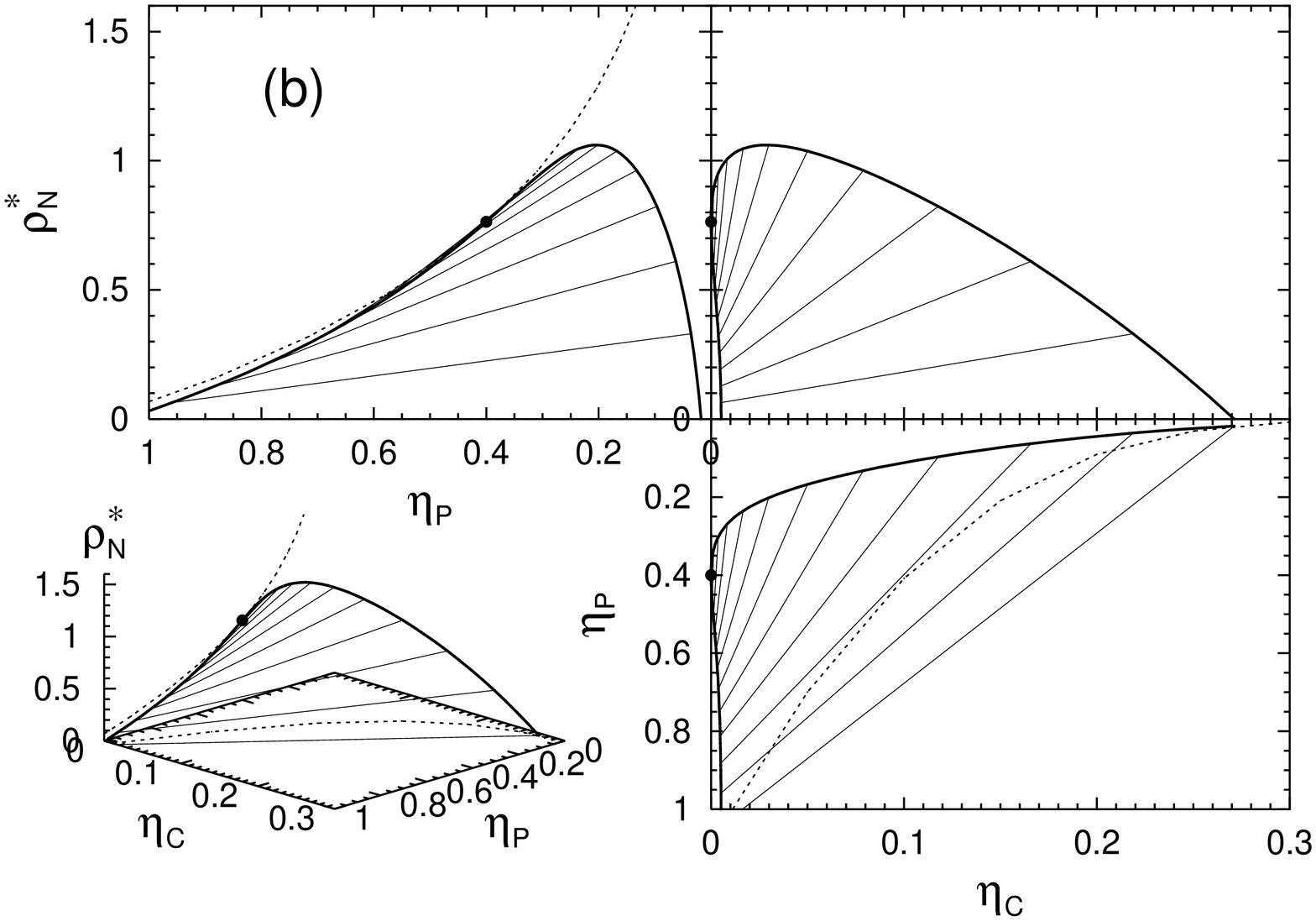}

    \includegraphics[width=\mypicwidth,angle=-90]{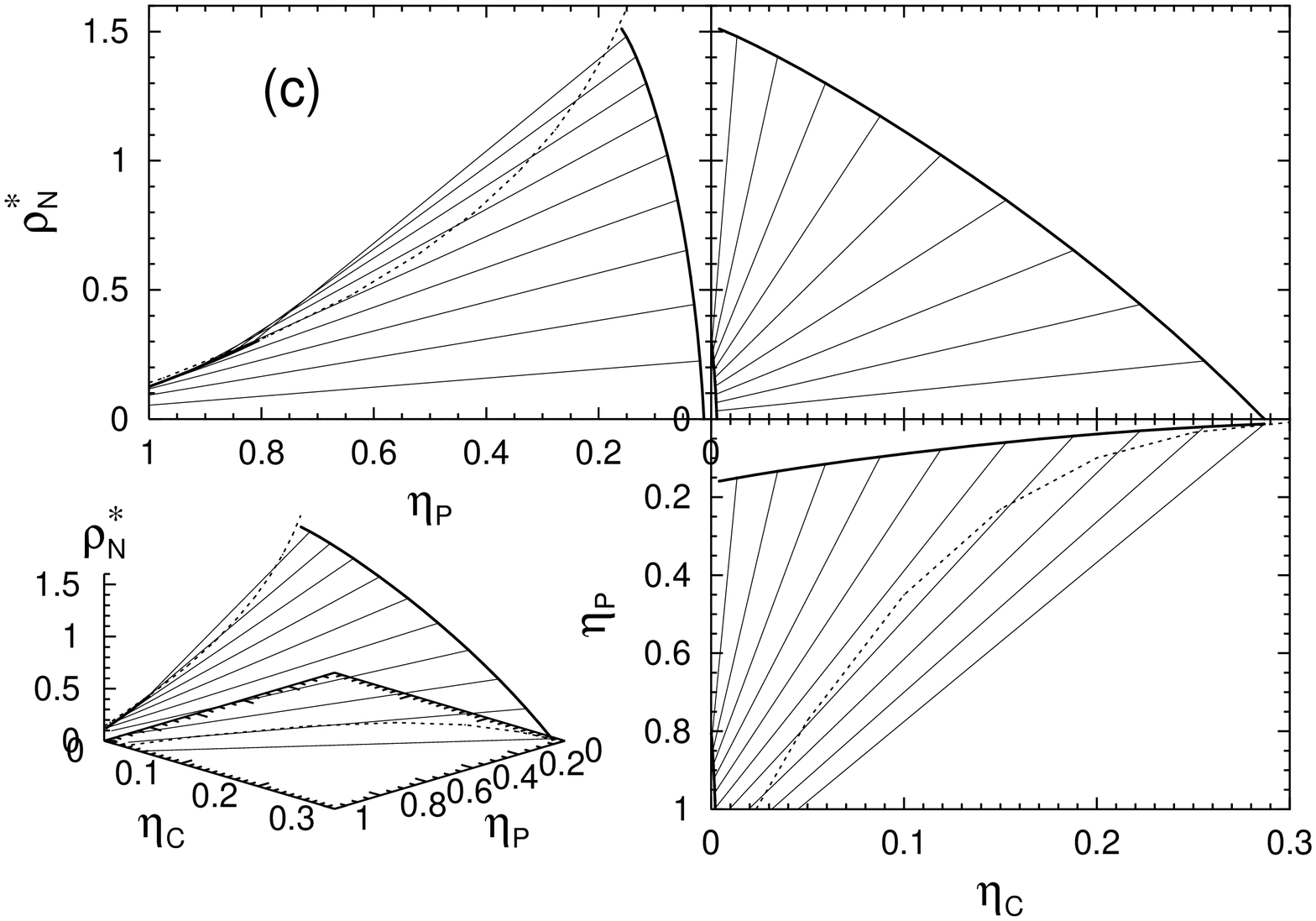}

    \caption{Demixing phase diagram of a ternary
     colloid-polymer-needle mixture with hard polymer-needle
     interactions for $\sigma_C=\sigma_P=L$, and $\eta_P^r=$0.8 (a),
     1.08731 (b), 1.2 (c).}  \label{FIGwr1} \end{center}
\end{figure}

We first consider equal particle sizes, $\sigma_C=\sigma_P=L$.  It
turns out that interesting behavior is observed only for small $\rn<
0.2$. The colloid-needle demixing curve lies well above this region,
and is only weakly affected by $\epr>0$.  In the absence of needles
($\rn=0$) and for large enough polymer density, colloids and polymers
demix, indicated by a miscibility gap along the $\rn=0$ axis (see
Fig.\ \ref{FIGwr1}a for $\epr=0.8$). Increasing needle density $\rn>0$
causes the gap to shrink and eventually to disappear in a critical
point. Quite surprisingly, and in contrast to the former case of
absent $PN$ interactions, the addition of needles {\em favors}
mixing. This behavior may reflect a competition between the depleting
effects of interacting polymers and needles.  By analogy with the $CP$
subsystem it is clear that at sufficiently high polymer density, a
$PN$ miscibility gap will open for $\ec=0$. However, this happens not
by growing a small bump as in the $CP$ case. Instead the $CP$ demixing
curve bends over to smaller $\ec$ and touches (with its critical
point) the $\ec=0$ axis (see Fig.\ \ref{FIGwr1}b for
$\epr=1.08731$). For larger $\epr$, the critical point disappears (see
Fig.\ \ref{FIGwr1}c for $\epr=1.2$).

\begin{figure}
  \begin{center}
    \includegraphics[width=\mypicwidth,angle=-90]{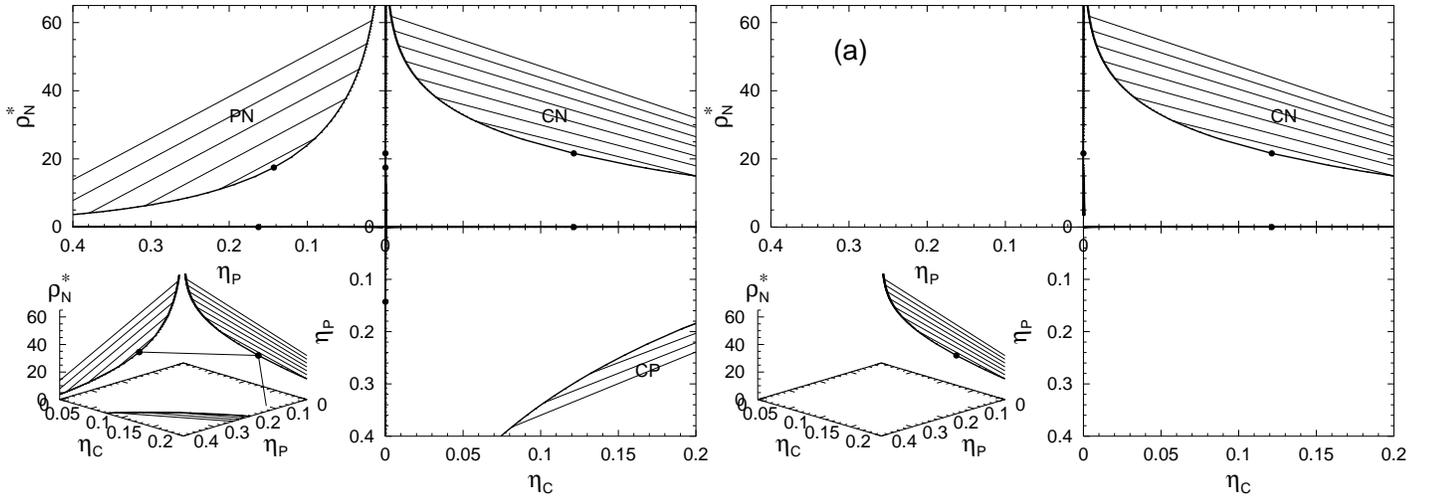}
    \caption{ Demixing phase diagrams in the binary subsystems with
    hard polymer-needle interactions for $\sigma_C=2\sigma_P=L/2$.}
    \label{FIGwr2two} \end{center}
\end{figure}

\begin{figure}
  \begin{center}

    \includegraphics[width=\mypicwidth,angle=-90]{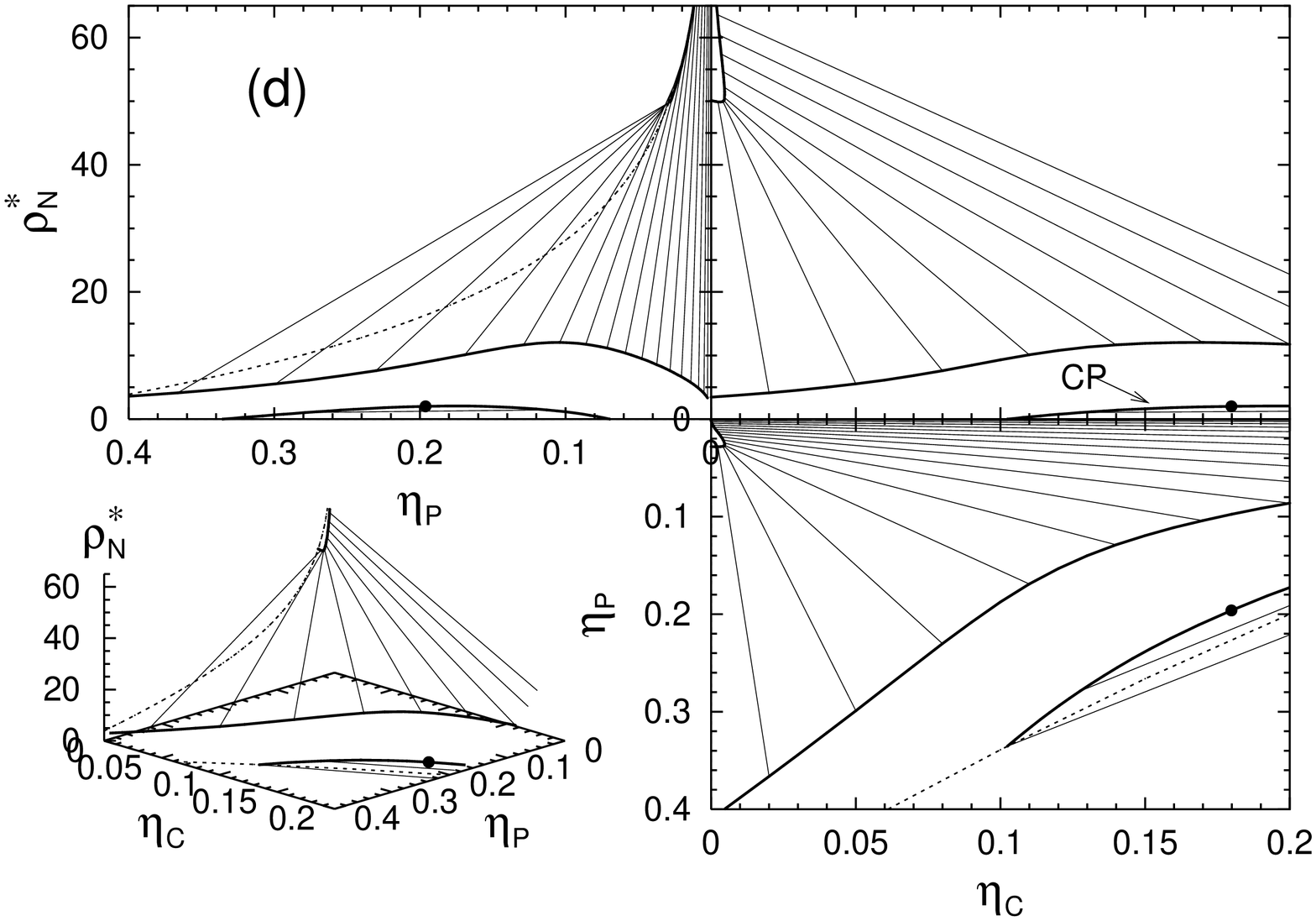}

    \includegraphics[width=\mypicwidth,angle=-90]{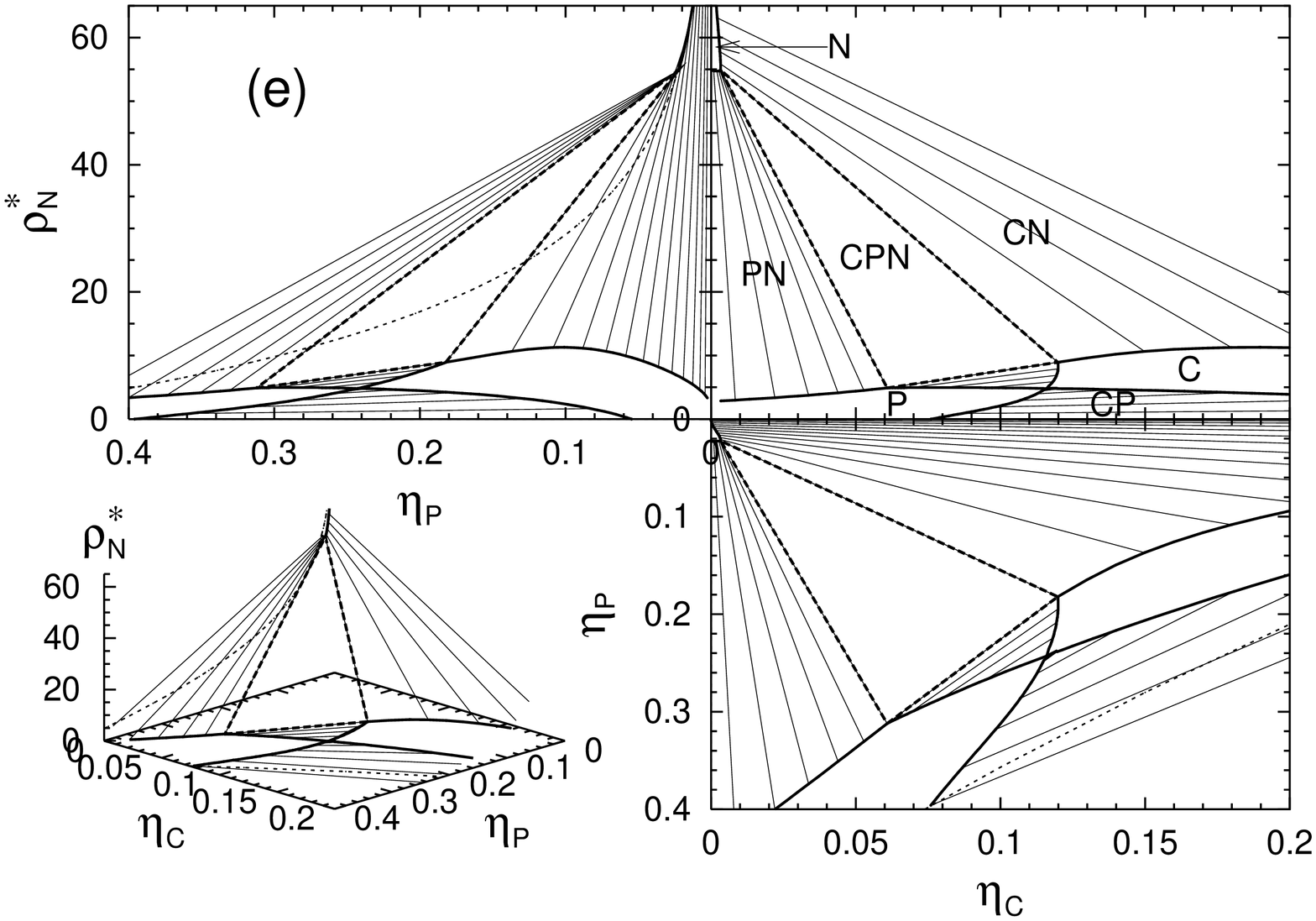}

    \includegraphics[width=\mypicwidth,angle=-90]{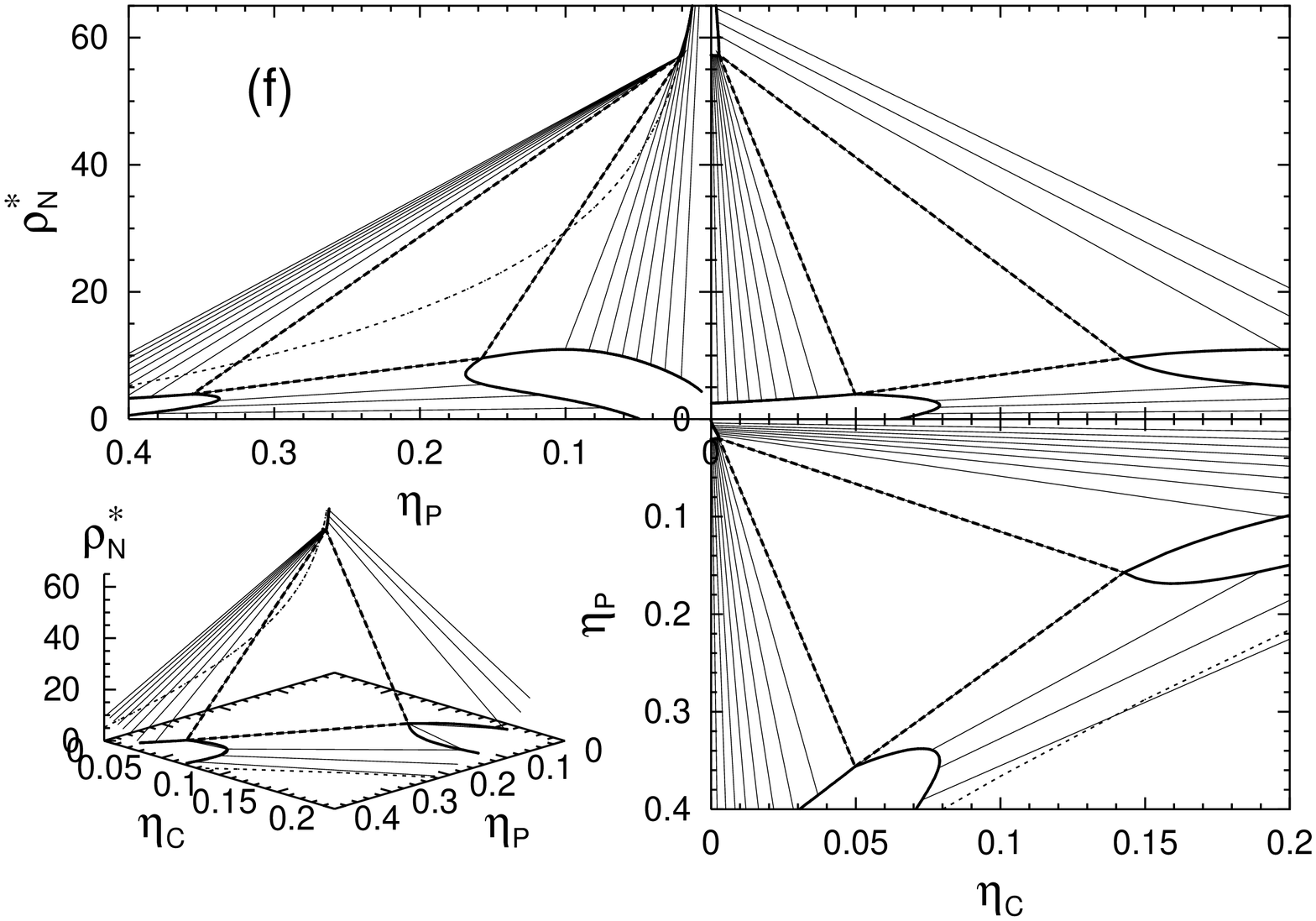}

    \caption{ Same as Fig.\ \ref{FIGwr1}, but for
    $\sigma_C=2\sigma_P=L/2$, and $\eta_P^r=$0 (a), 0.4 (b), 0.408107
    (c), 0.5 (d), 0.52626 (e), 0.54 (f).}  \label{FIGwr2} \end{center}
\end{figure}

In order to bring $CP$ and $CN$ demixing closer together, we consider
a reduced polymer size $\sigma_P=\sigma_C/2$, generating a weaker
depletion attraction between colloids (at the same number density of
polymers), and longer needles, $L=2\sigma_C$ generating stronger
depletion between colloids, and hence lower $\rn$ at the critical
point in the binary $CN$ case. Figure \ref{FIGwr2two} shows the
binodals in the (three) binary subsystems.  For the ternary mixture,
we follow a path of increasing $\epr$, starting with $\epr=0$, for
which the phase diagram is displayed in Fig.\ \ref{FIGwr2}a. There is
no polymer present in the system, and phase separation into
colloid-rich and needle-rich phases occurs at high enough densities of
these components. Both $\ep-\rn$ and $\ec-\ep$ planes are inaccessible
as $\ep=0$. Increasing polymer density ($\epr=0.4$ in
Fig.\ref{FIGwr2}b) shifts the $CN$ critical point to lower $\ec$,
distorting the formerly rounded shape of the binodal. For $\ec=0$,
polymers and needles demix, as $\epr$ is above the critical value for
the Widom-Rowlinson type demixing of these species. The presence of
colloids ($\ec>0$) disturbs the $PN$-transition; the miscibility gap
narrows, eventually disappearing in a critical point, with subsequent
miscibility. At $\epr=0.408107$ (Fig.\ \ref{FIGwr2}c) the $CN$ and
$PN$ critical points merge into a single one, and a needle-rich phase
($N$) becomes isolated. This coexists with a phase that consists
(primarily) of colloids and polymers at varying composition.  For
growing $\epr$, the ``double'' critical point broadens into a line and
results in a thin neck joining both transitions.

With increasing $\epr$ the coexistence region broadens further (see
Fig.\ \ref{FIGwr2}d for $\epr=0.5$). Colloids and polymers also
demix. For $\rn=0$, the system is above the critical point for the
pure AO model, and hence coexistence between colloid-rich and
polymer-rich phases occurs. Again, the presence of the third
component, in this case $N$, causes the density gap to shrink and
eventually disappear with a critical point. As all binary subsystems
are by now demixed, it is evident that the system will ultimately
display coexistence between three phases, each one enriched by one of
the components, and represented by a triangle in system
representation. Each corner of the triangle corresponds to one of the
three coexisting phases.  The Gibbs phase rule dictates that one
degree of freedom remains, which is $p_{\rm tot}$ or, equivalently,
$\epr$ (note that for hard-body systems, temperature is trivially
related to pressure).  It is striking, however, how this triangle
develops. One might expect this to occur by the joining of existing
binary coexistence regions. This is not the case. The ternary region
instead grows solely out of the $N$-rich--poor coexistence, whereby
$CP$-coexistence is only a spectator, separated by mixed states. The
initial three-phase triangle is extremely elongated (being a line as a
boundary case). One corner corresponds to a needle-rich phase; both
others differ only slightly in densities, one phase favoring colloids,
the other polymers. Moving away from this $CP$-edge of the triangle
(by reducing $\rn$) leads to binary coexistence between $C$ and
$P$. This phase separation is reminiscent of the behavior of the pure
AO model.  However this reentrant coexistence is triggered by the
presence of the needles, and it is separated (by mixed states) from
the pure AO transition (and its region of stability in the presence of
needles).  In Fig.\ \ref{FIGwr2}e we show results for $\epr=0.52626$,
where the critical points of both $CP$ transitions have already
merged, and again a neck is reminiscent of the formerly distinct
transitions. For still larger $\epr$, the three-phase triangle grows
further (see Fig.\ \ref{FIGwr2}f for $\epr=0.54$).  Ultimately, at
sufficient concentration the colloids must freeze, but we disregard
the solid phase in the present work.  We finally note that the whole
scenario is covered over a relatively small density interval
$\epr=0.4-0.54$, and that the packing fractions of colloids and
polymers are only moderate.  However, needle densities can be quite
high.

\section{Discussion}
\label{SECdiscussion}
In conclusion, we have considered a simple hard-body model for a
mixture of spherical colloidal particles, globular polymer coils and
needle-shaped objects, which may represent either colloidal needles,
stretched polymers or polyelectrolytes. We have extended a recent DFT
approach to this model and applied it to bulk fluid phases.  The
resulting phase behavior is very rich, ensuing from competition of
demixing in the binary subsystems.

The present work has interesting implications for the techniques of
integrating out degrees of freedom (see e.g.\
\cite{brader00,brader01inhom}). Note that by integrating out, e.g.,
the needles, effective interactions between pairs of colloids, pairs
of polymers, as well as colloids and polymers arise. Hence one arives
at a binary mixture with (soft) depletion interactions. To what extent
the ultimate mapping onto a one-component (colloid) system, by further
integrating out the polymers, can be achieved is an interesting
question.  As a further outlook, the inclusion of freezing of
colloids, disregarded in the present work, would further enrich the
phase behavior. Computer simulations are desirable to test the
theoretical phase diagrams. Furthermore it is interesting to elucidate
the structural correlations present in the various fluid phases.
Inhomogeneous situations, such as induced by walls or present at
interfaces between demixed states, constitute further exciting
directions of research.

\vspace{10mm} We acknowledge useful discussions with Stuart G.\ Croll.

*Permanent address: Institut f{\"u}r Theoretische Physik II,
  Heinrich-Heine-Universit{\"a}t D{\"u}sseldorf,
  Universit{\"a}tsstra{\ss}e 1, D-40225 D{\"u}sseldorf, Germany.


\begin{figure}
  \begin{center}

    \includegraphics[width=\mypicwidth,angle=-90]{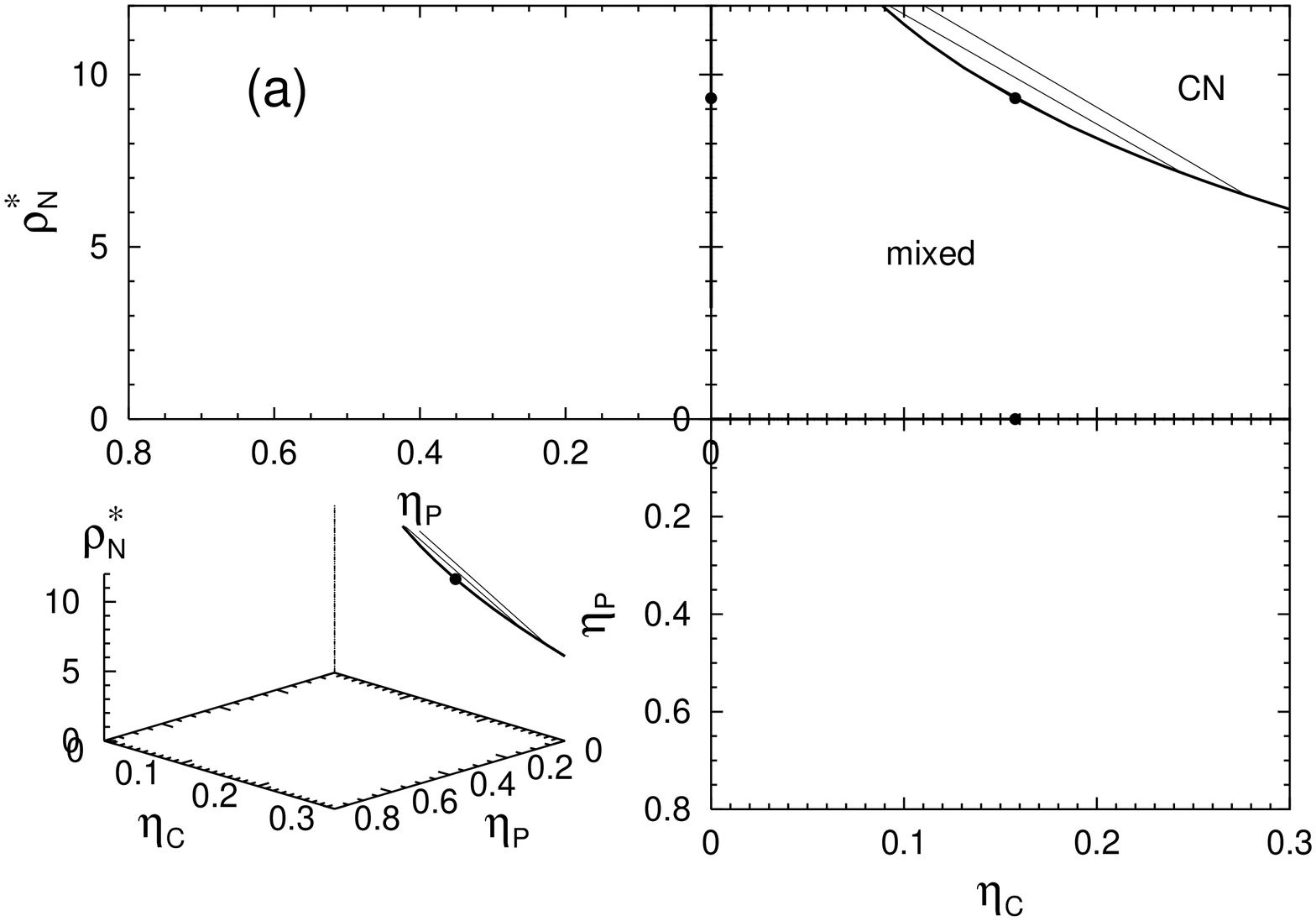}

    \includegraphics[width=\mypicwidth,angle=-90]{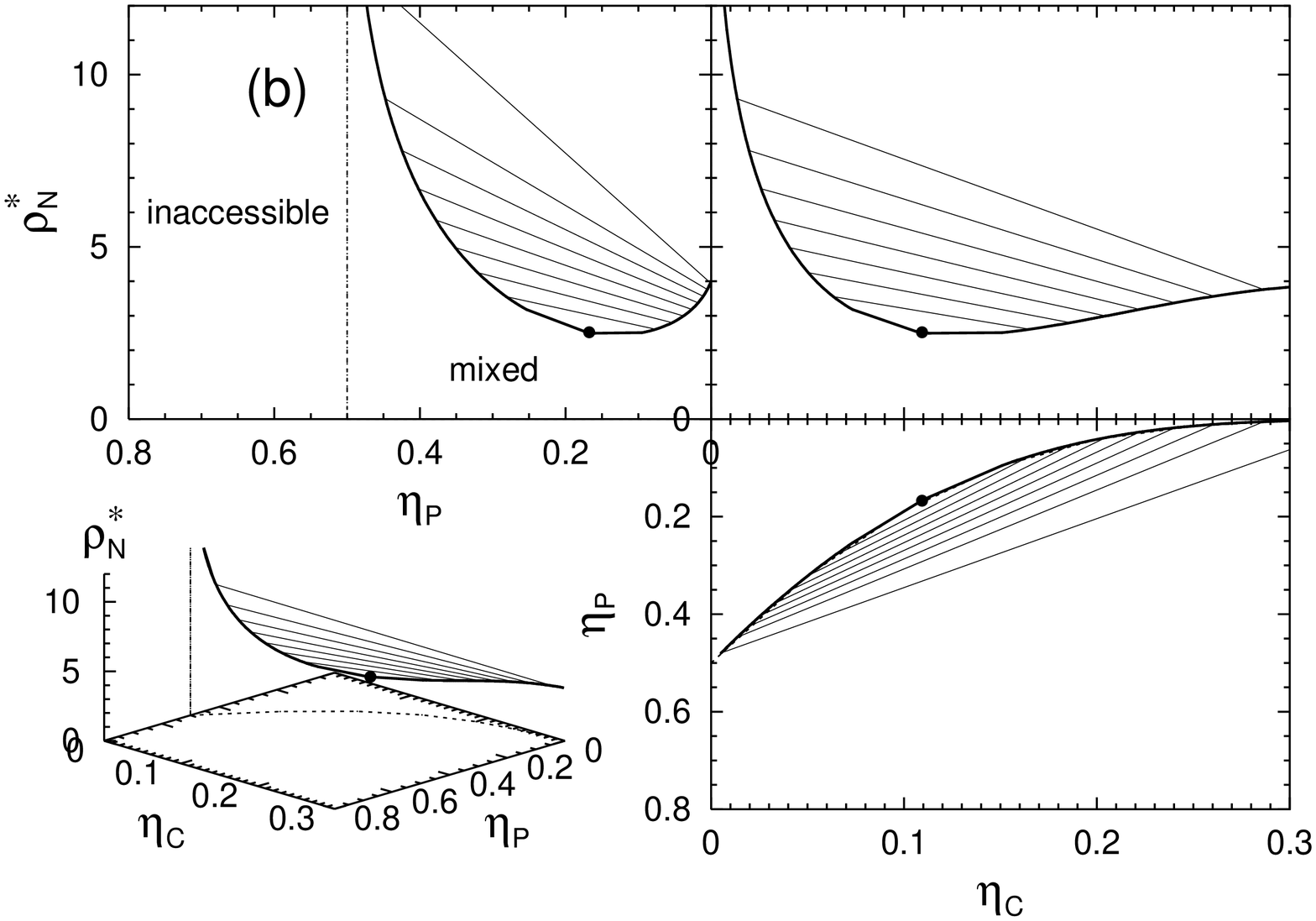}

    \includegraphics[width=\mypicwidth,angle=-90]{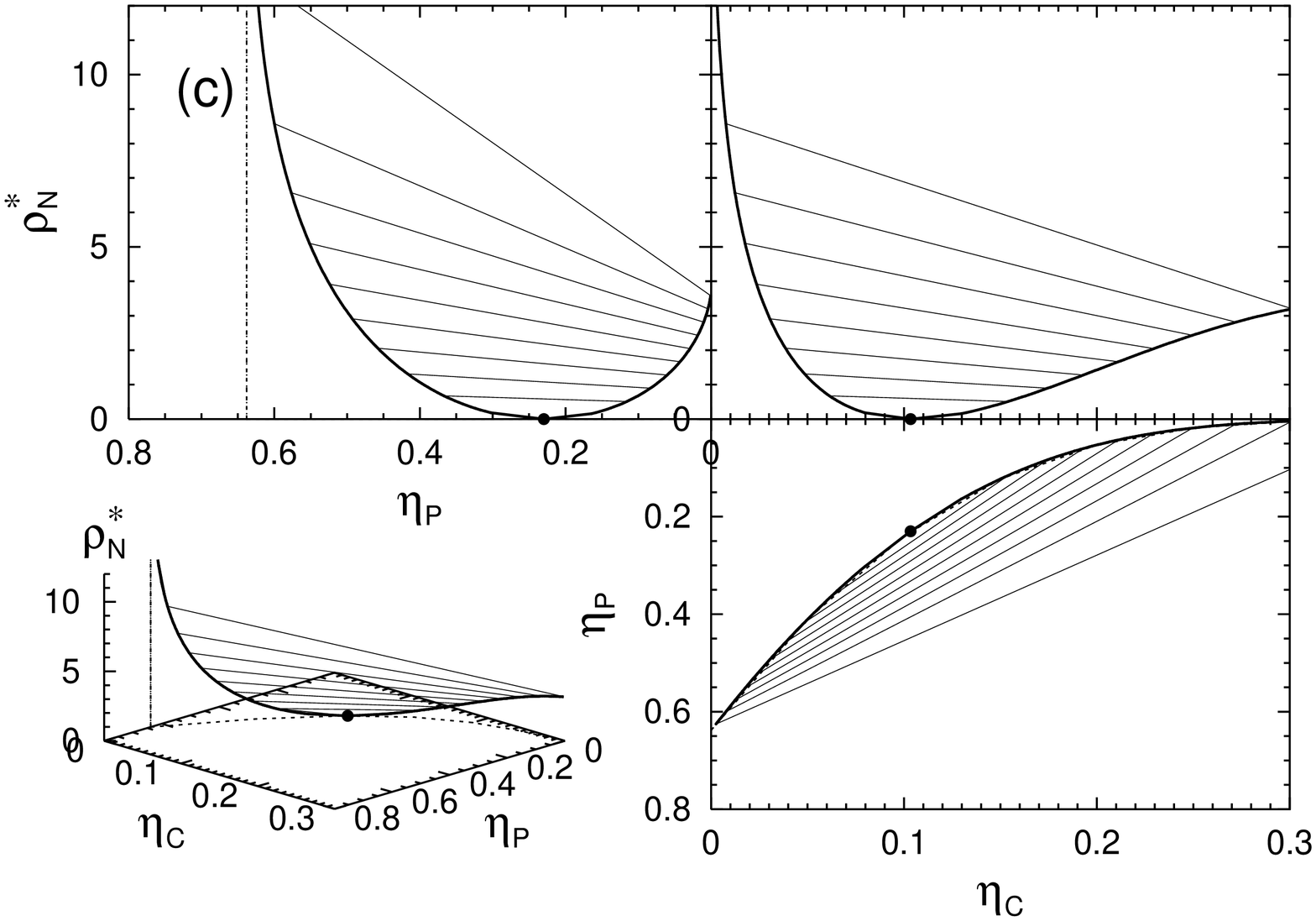}

    \end{center}
\end{figure}

\begin{figure}
  \begin{center}

    \includegraphics[width=\mypicwidth,angle=-90]{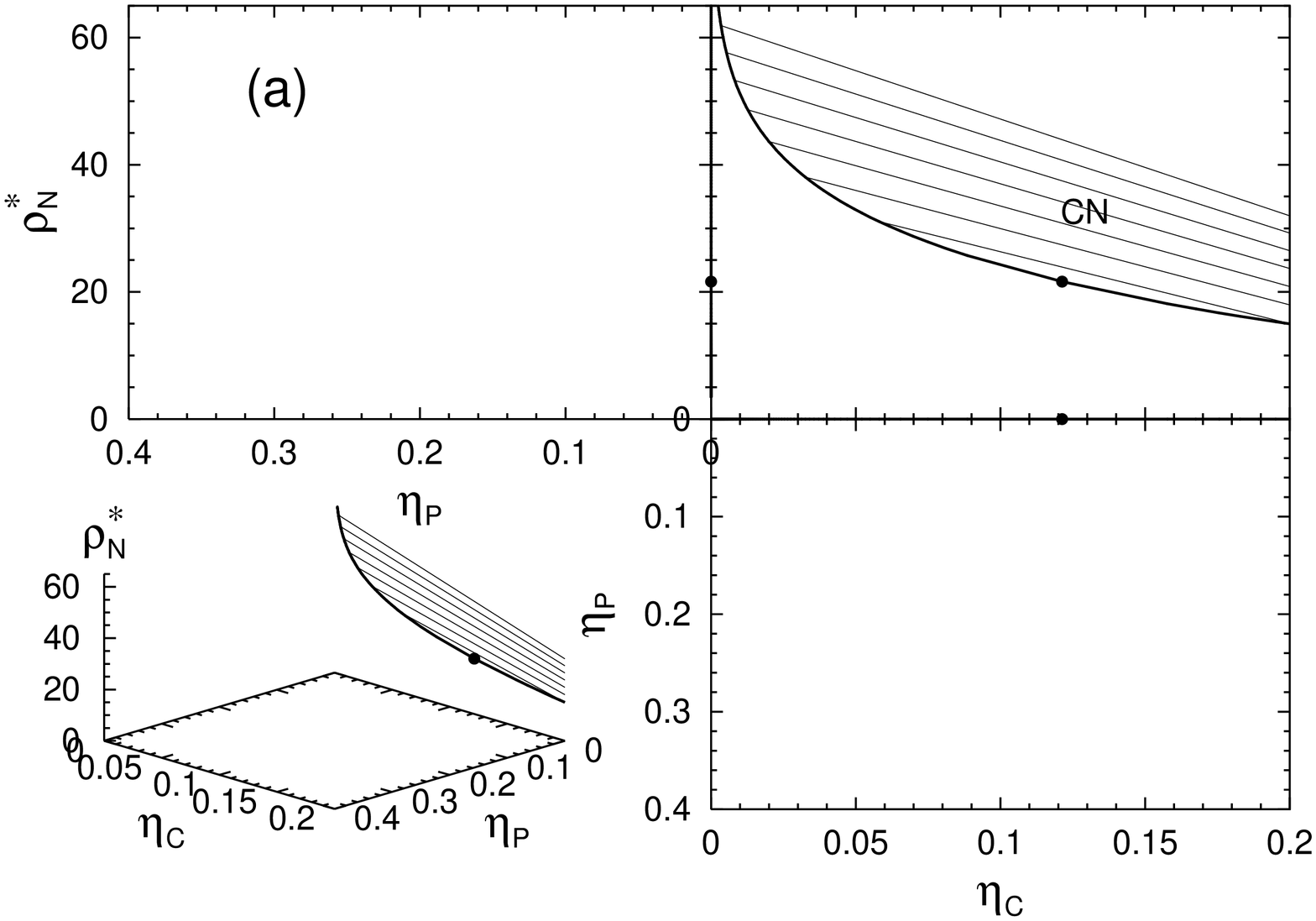}

    \includegraphics[width=\mypicwidth,angle=-90]{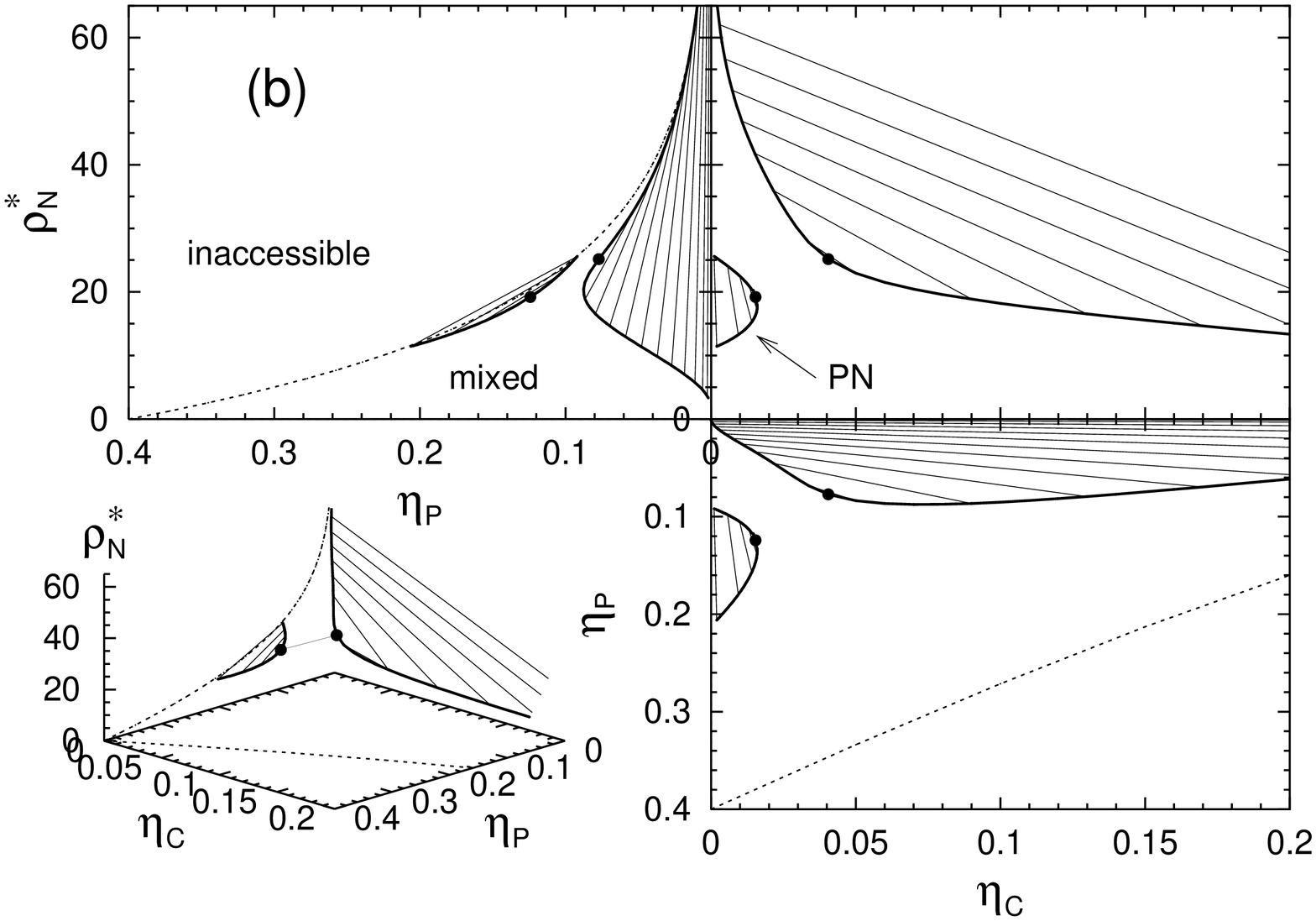}

    \includegraphics[width=\mypicwidth,angle=-90]{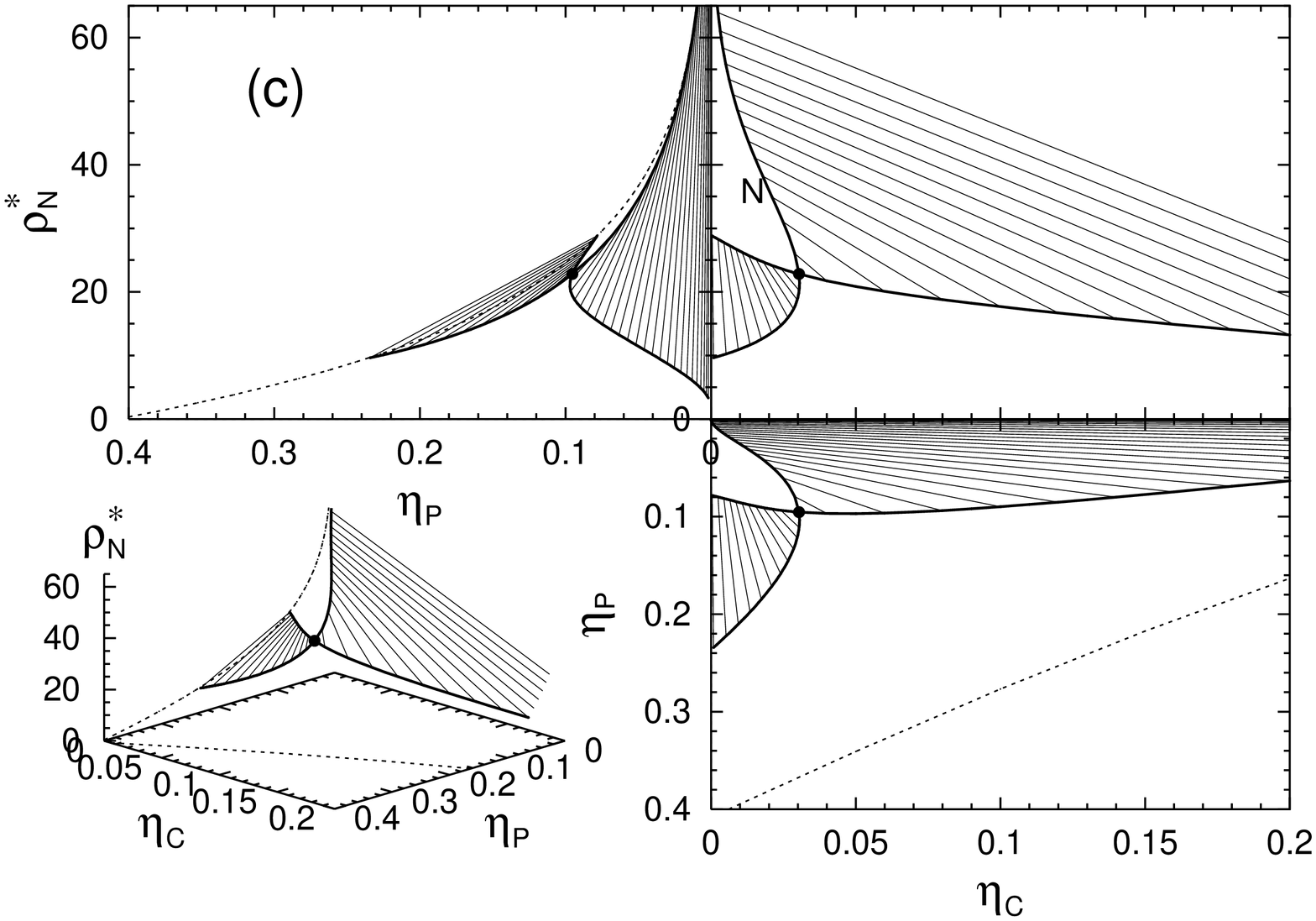}

  \end{center}
\end{figure}

\clearpage

\end{document}